**Machine Learning Materials Properties with Accurate Predictions, Uncertainty Estimates, Domain Guidance, and Persistent Online Accessibility**

**Authors:** Ryan Jacobs[1], Lane E. Schultz[1], Aristana Scourtas[2,3], KJ Schmidt[2,3], Owen Price-Skelly[2,3], Will Engler[2,3], Ian Foster[3,4], Ben Blaiszik[2,3], Paul M. Voyles[1], Dane Morgan[1]

[1] Department of Materials Science and Engineering, University of Wisconsin-Madison, Madison, WI, USA.

[2] Globus, University of Chicago, Chicago, IL, USA.

[3] Data Science and Learning Division, Argonne National Laboratory, Lemont, IL, USA.

[4] Department of Computer Science, University of Chicago, Chicago, IL, USA

**Abstract**

One compelling vision of the future of materials discovery and design involves the use of machine learning (ML) models to predict materials properties and then rapidly find materials tailored for specific applications. However, realizing this vision requires both providing detailed uncertainty quantification (model prediction errors and domain of applicability) and making models readily usable. At present, it is common practice in the community to assess ML model performance only in terms of prediction accuracy (e.g., mean absolute error), while neglecting detailed uncertainty quantification and robust model accessibility and usability. Here, we demonstrate a practical method for realizing both uncertainty and accessibility features with a large set of models. We develop random forest ML models for 33 materials properties spanning an array of data sources (computational and experimental) and property types (electrical, mechanical, thermodynamic, etc.). All models have calibrated ensemble error bars to quantify prediction uncertainty and domain of applicability guidance enabled by kernel-density-estimate-based feature distance measures. All data and models are publicly hosted on the Garden-AI infrastructure, which provides an easy-to-use, persistent interface for model dissemination that permits models to be invoked with only a few lines of Python code. We demonstrate the power of this approach by using our models to conduct a fully ML-based materials discovery exercise to search for new stable, highly active perovskite oxide catalyst materials.

# 1. Introduction

Data-centric machine learning (ML) and artificial intelligence (AI) approaches have witnessed rapid and widespread adoption in materials science and engineering over the past decade.[1–10] ML and AI tools are being applied widely in the discovery, design, and understanding of materials, for example to predict materials properties,[11–16] perform automated computer vision analysis of microscopy images,[17–23] construct ML-based interatomic potentials,[24–31] extract and curate data and knowledge using natural language processing[32–36] and large language models,[37–40] use of generative models to propose novel crystal structures and chemistries for inverse materials design,[41–44] and combining one or more of the above approaches to formulate fully autonomous self-driving laboratories.[45–52] Here, we focus on using ML to predict materials properties for materials discovery and design.

ML models can generate predictions of materials properties orders-of-magnitude faster than resource-intensive simulations or experiments. Recent studies of ML predictions of materials properties have grown the ability to use ML for materials exploration via a range of approaches. Researchers have, for example, performed fits to many properties together,[12,53,54] employed novel approaches to featurization,[55,56] introduced new model types,[53] and integrated ML property models with follow-on materials screening and design workflows.[55,57,58] Regarding fits to many properties, Wang et al.[53] fit models to 28 materials properties, while Dunn et al.[12] fit to 13 materials properties and formulated the MatBench[59] leaderboard of materials property predictions. In addition, Choudhary et al.[54] formulated the Joint Automated Repository for Various Integrated Simulations for Machine Learning (JARVIS-ML), which reported fits to about 25 calculated materials properties and developed web applications for users to easily make new predictions. They also created the JARVIS Leaderboard of 274 benchmarks across 152 methods for materials in application areas of AI, electronic structure, force-fields, quantum computation, and experiments.[60] Regarding new descriptors, Zhai et al.[55] introduced a novel Lewis acid descriptor to improve fits to perovskite area specific resistance (ASR) values, which enabled the discovery of the new perovskite catalyst $Sr_{0.9}Cs_{0.1}Co_{0.9}Nb_{0.1}O_3$ with state-of-the-art high activity for the oxygen reduction reaction. Schindler et al.[56] devised a novel featurization scheme to distinguish between different crystallographic orientations and terminations of materials, enabling the construction of a random forest (RF) ML model able to accurately predict the work function, a highly surface-sensitive electronic property. Regarding the creation of new models, Wang et al.[53] devised the compositionally restricted attention-based network (CrabNet) model, which uses only composition-based input to generate features for an attention-based deep neural network that outperforms RF models for numerous materials properties, particularly if the datasets were large. Regarding ML models used for follow-on materials discovery, Zhai et al.[55] and Jacobs et al.[57] built ML models to use for screening of new perovskite catalyst materials, while Lu et al.[58] screened several promising materials for photovoltaic applications by first applying an ML model for bandgaps and then employing targeted density functional theory (DFT) calculations to validate the ML results. Beyond such single-pass screening to discover new materials for a specific application from an initial, pre-defined list of candidates, materials property models are also commonly employed in active learning-based materials discovery campaigns where the ML model is improved on the fly and integrated with higher fidelity computation or experiment to iteratively search for new materials in a prescribed design space.[61–69]

Yet despite these many successes of ML materials property models for materials discovery and design, critical limitations remain, which we address in this work and highlight in the ML materials property model development cycle shown in **Figure 1**. The first limitation relates to uncertainty quantification, which is absent in nearly all examples mentioned above. It is common practice in the ML model fitting stage of the development cycle to focus on finding a particular model and feature set that minimizes a standard statistical metric, for example the root-mean-square-error (RMSE) from 5-fold cross validation. We argue that this is a necessary, but not sufficient, criterion for development of a useful ML model for materials property prediction. We contend that a useful materials property ML model should satisfy three criteria: (i) *accurate prediction*, meaning that the model has low residuals and therefore a low RMSE (e.g., 5-fold cross validation), (ii) *error quantification*, meaning it has error estimates that well-represent the true error of a prediction, and (iii) *domain quantification*, meaning it has estimates for the likelihood that the user is making predictions outside the domain of applicability of the model. From the authors' collective knowledge of the literature and experience in the field, it is rare that a model is provided with reliable error and domain quantification. This is likely because methods for obtaining reliable error estimates[70–76] and assessing domain of applicability[77–83] are both active areas of continuing research, and no single approach has emerged and been adopted as standard practice for materials property prediction.

The second limitation relates to data and model accessibility. Data accessibility has received a lot of attention in recent years, and, as of this writing, excellent guidance on best practices as well as accessible and easy-to-use repositories are available.[12,59,84–86] However, it can still be difficult to obtain a dataset in a form ready for ML applications, which is important for enabling quick and reliable reproduction of previous work and easy integration into improved or new model fits. For example, data is often supplied without the final features used, or sufficient information or working software to enable generation of those features. One approach to avoid this problem is to share data in an ML-ready form, meaning the data has all necessary input and output and metadata for immediate ML model use and assessment. All datasets used in this work are hosted on Foundry-ML.[84] Briefly, Foundry-ML is a software platform designed to make datasets easily accessible for training and reproducing ML model performance. It also provides users the ability to publish and search for ML-ready datasets in a variety of scientific disciplines.

While model accessibility has received less attention, it is a major issue hampering reproducibility and reuse of ML models outside of the researcher(s) who developed them. It is common for researchers to share a fitted model and the code to run the model on open software repositories like GitHub. While helpful, the useability of such models tends to degrade in the medium-to-long term (e.g., greater than a year or two after initial publication, and sometimes sooner). The main reasons for model useability to degrade over time are updates and differences in computer operating systems, Python versions, software package dependencies, and more, where conflicts or completely different outputs may be obtained with different software versions or operating systems. Overcoming these challenges can be time-consuming, but are necessary to resolve to get the code working, hindering reproducibility. One approach to ensure model persistence is by using a containerization service, e.g., Docker. A drawback to services like Docker is that the resulting Docker image may not be straightforward for users to access or execute, creating barriers to ease of use.

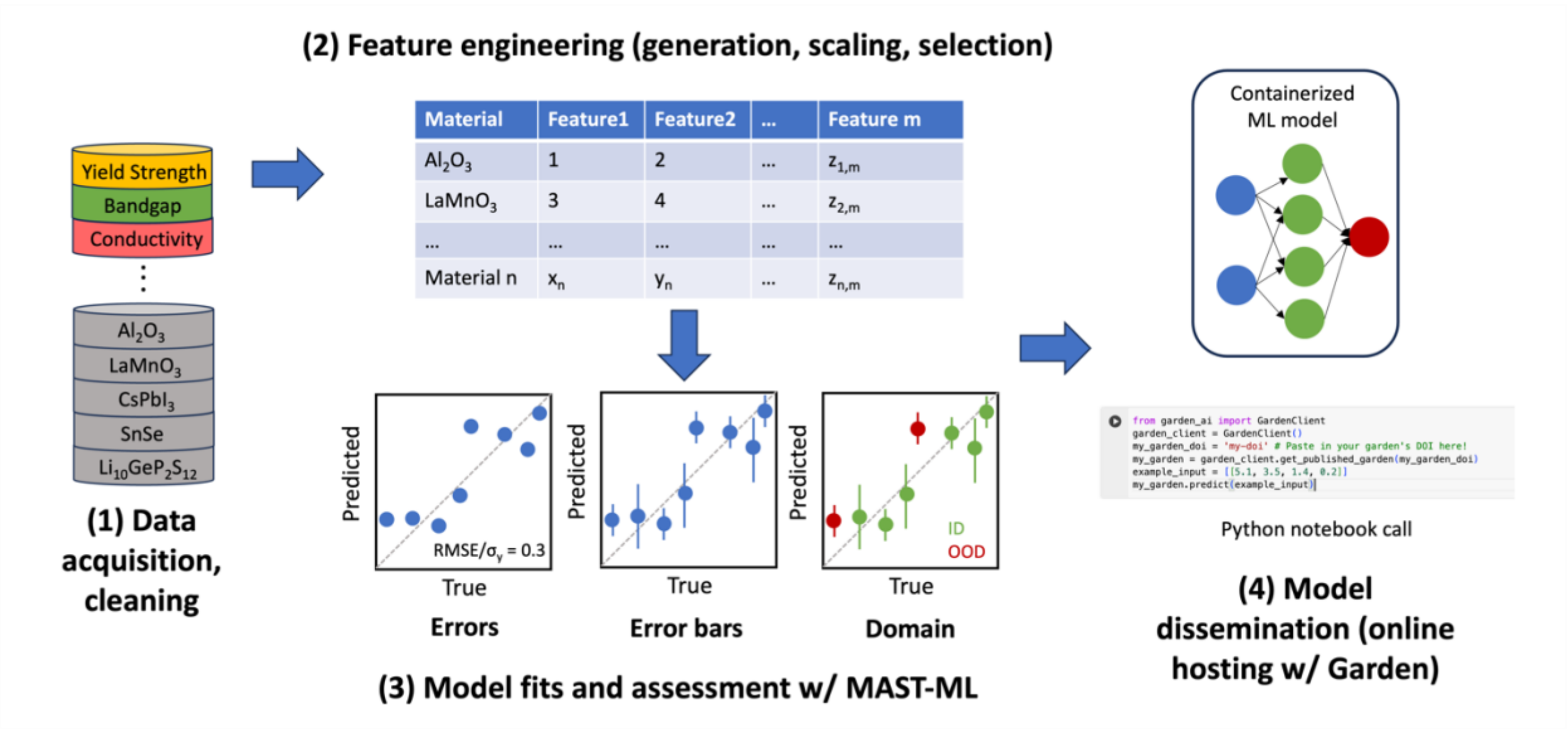

**Figure 1.** An overview of the ML production workflow for generating ML models of materials properties used in this work.

In this work, we demonstrate an approach that dramatically improves the uncertainty quantification and produce readily usable ML models of materials properties. We accomplish this in two ways. First, we improve upon the typical approach of ML model fitting to automatically report reliable error bar uncertainty estimates and measures of model domain of applicability in addition to providing robust prediction statistics. This improvement in ML model fitting is accomplished using our MAterials Simulation Toolkit for Machine Learning (MAST-ML)[87] package. Second, we introduce the Garden-AI software infrastructure,[88] which is an ML model hosting platform which creates readily useable, persistent, and citable containers of fit ML models, enabling running of an ML model with only a couple lines of Python code on distributed computing endpoints.[89] We demonstrate our overall approach by fitting ML models to 33 distinct materials properties, complete with error bars and domain estimates, and hosting all 33 of these models on Garden-AI. We then illustrate the power of our approach by using a subset of our models to quickly screen for perovskite catalytic materials predicted to be promising for fuel cell and electrolyzer applications. Taken together, we believe the union of accessible materials data, informative ML models with error bars and domain of applicability guidance, and model accessibility via the Garden-AI infrastructure can increase the use and impact of ML for property prediction in materials discovery and design.

# 2. Data and Methods

## 2.1. Materials datasets

We retrieved from the literature the 33 materials property datasets listed in **Table 1**. This collection of datasets is by no means exhaustive. Rather, it is meant to demonstrate the approach and to provide a seed upon which the community can build. We sought in assembling this collection to include a wide array of both materials chemistries and property types, with the latter encompassing electronic (e.g., bandgap), mechanical (e.g., yield strength), thermodynamic/catalytic (e.g., material stability represented as convex hull energy), and kinetic (e.g., thermal conductivity). **Table 1** provides for each dataset the materials property, dataset size, data and property type, and data source. Broadly, the data consist of both experimentally measured properties and calculated values (all from DFT simulations), and the dataset sizes range from small (137 points for perovskite thermal expansion coefficient) to modest in size (9646 points for perovskite formation energies). While references to the original data are given for all properties, we reiterate that all of the datasets used in ML model fitting are also hosted on Foundry-ML in a ready-to-use programmatically accessible format.[84]

**Table 1.** Summary of the 33 materials property datasets investigated in this work. Property names are listed in alphabetical order. Abbreviations: $R_c$ = critical cooling rate, $D_{max}$ = maximum cast diameter, ASR = area specific resistance. TEC = thermal expansion coefficient, RPV = reactor pressure vessel, $T_c$ = superconducting critical temperature.

| Property name | Data type | Property type | Number of data points | Data reference |
|---|---|---|---|---|
| Bandgap | Experiment | Electronic | 6354 (6031 after cleaning)* | 90 |
| Concrete compressive strength | Experiment | Mechanical | 1030 | 91 |
| Debye Temperature | Computed | Thermodynamic / catalytic | 4896 | 92 |
| Dielectric constant | Computed | Electronic | 1056** | 93 |
| Dilute solute diffusion | Computed | Kinetic | 408 | 94 |
| Double perovskite bandgap | Computed | Electronic | 1306 | 95 |
| Elastic tensor (bulk modulus) | Computed | Mechanical | 1181** | 96 |
| Exfoliation energy | Computed | Thermodynamic / catalytic | 636 | 54 |
| Heusler magnetization | Computed | Electronic | 370 | 97 |
| High entropy alloy hardness | Experiment | Mechanical | 370 | 65 |
| Lithium conductivity | Experiment | Kinetic | 372 | 98 |
| Metallic glass $D_{max}$ | Experiment | Kinetic | 998 | 99 |
| Metallic glass $R_c$ (from LLM) | Experiment | Kinetic | 297 | 100 |
| Metallic glass $R_c$ | Experiment | Kinetic | 2125 | 101 |
| Mg alloy yield strength | Experiment | Mechanical | 365 | 102 |
| Oxide vacancy formation | Computed | Thermodynamic / catalytic | 4914 | 103 |
| Perovskite ASR | Experiment | Thermodynamic / catalytic | 289 | 104 |
| Perovskite conductivity | Experiment | Kinetic | 7230 | 105 |
| Perovskite formation energy | Computed | Thermodynamic / catalytic | 18,928 (9646 after cleaning many duplicate compositions)*** | 106 |
| Perovskite H absorption | Experiment | Thermodynamic / catalytic | 795 | 107 |
| Perovskite O p-band center | Computed | Electronic | 2912 | 108 |
| Perovskite stability | Computed | Thermodynamic / catalytic | 2912 | 108 |
| Perovskite TEC | Experimental | Mechanical | 137 | 109 |
| Perovskite work function | Computed | Electronic | 613 | 110 |
| Phonon frequency | Computed | Mechanical | 1265 | 111 |
| Piezoelectric max displacement | Computed | Electronic | 941 | 112 |
| RPV transition temperature shift | Experiment | Mechanical | 4535 | 113 |
| Semiconductor defect levels | Computed | Electronic | 896 | 114 |
| Steel yield strength | Experiment | Mechanical | 312 | 115 |
| Superconductivity $T_c$ | Experiment | Electronic | 6252 | 116 |
| Thermal conductivity | Experiment | Kinetic | 872 | n/a |
| Thermal conductivity (AFLOW) | Computed | Kinetic | 4887 | 92 |
| Thermal expansion | Computed | Mechanical | 4886 | 92 |

*Compositions were removed which contained characters not interpretable as a pymatgen composition.

**A much larger database is available on Materials Project; this data was from the original paper.

***For materials with duplicate compositions, the material with lower formation energy (more stable) was kept.

## 2.2. Machine learning model fits

All ML model fits and analysis were conducted using the MAST-ML package.[87] Briefly, MAST-ML is an open-source software package designed to accelerate and streamline the production of ML models, with a particular focus on regression models for predicting materials properties. The initial version of MAST-ML was published in 2020.[87] Since then, the code has been upgraded to include state-of-the-art approaches for model uncertainty quantification and updated to enhance ease-of-use. The uncertainty quantification capabilities include both assessment of model error bars and model domain of applicability, explained more in **Section 2.3** and **Section 2.4**, respectively. RF models were used to fit each of the 33 materials property datasets described in **Section 2.1**. RF was chosen as it was found to predict each property well, with error values generally in good agreement with literature-reported values from other studies, some of which also used RF but many of which used other model types. In addition, RFs tend to be easy to fit from a practical standpoint, as they require minimal hyperparameter tuning. In this work, the only parameter which varied between datasets was the number of decision tree estimators in the RF. Finally, RFs are ensembles of decision trees, which makes them particularly amenable to our approach of quantifying error bars, which we describe in more detail in **Section 2.3**. In general, the most popular ML models in materials property prediction (as of early 2024) are ensemble models such as RF, XGBoost, and gradient boosting models, with RF emerging as the most popular by a wide margin.[10]

The features used to characterize material compositions to fit the ML models belong to one of three groups: (i) The first class of features is mathematical combinations of elemental properties (e.g., atomic radius, electronegativity, whether element is a rare earth, etc.) to describe the composition (e.g., for $Al_2O_3$, properties may be composition-averaged atomic radius of Al and O). This approach of using elemental features has seen widespread use, and MAST-ML enables featurization based on key elemental property sets from our own work,[117] from the MAGPIE approach of Ward et al.,[11,118] and from the composition-based feature vector (CBFV) approach.[119] (ii) The second class of features is categorical features which are turned into numerical vectors via one-hot encoding. For example, in the Heusler magnetization dataset, categorical information regarding the Heusler structure type is given. Including the one-hot encoding of Heusler type was found to significantly improve the ML model for this property. (iii) The final class of features is what we call material-specific features. These features are not general like the elemental features described above but are critical features to describe correlations with the target variable within a specific dataset or class of materials. For example, for the concrete compressive strength dataset, some features are the amount of water in the concrete mixture, the amount of coarse aggregate, etc. As another example for the RPV transition temperature shift dataset, material irradiation conditions such as the flux and fluence (total dose) of irradiation are key factors governing the transition temperature shift and are unique to that particular dataset. For the models using the elemental-based features described above, the as-made feature set typically contains several hundred features. Based on the authors' collective experience in the field, for most material property datasets of modest size (e.g., a couple hundred to a couple thousand data points), 25-50 elemental features tend to be sufficient to give fits of good statistical quality without significant overfitting, at least with RF models. Therefore, for those properties using only elemental properties, we default to using 25 features, and for a couple cases of larger datasets, 50 features were used. Features were selected using the RF feature importance rankings.

Some of the datasets explored in this work contain full atomic structure information (e.g., dielectric constant, elastic tensor, piezoelectric max displacement datasets). Given this detailed structural information, it is very likely that fits with lower errors could be obtained using state-of-the-art structure-aware ML models like the MatErials Graph Network (MEGnet)[120] or the Atomistic Line Graph Neural Network (ALIGNN).[121] However, we restrict our use-case to features obtained solely from material compositions, categorical features, or materials-specific features as described above, as our domain of applicability approach has only been validated using these types of features. We note that our ML error bar approaches are amenable to the use of structural features, provided such features can be represented in a tabular format. The use of GNN-based ML models for representing structure have not yet been validated for use with our error bar and domain approaches, and doing so is beyond the scope of this work. Deep learning neural networks like ElemNet have been developed to work solely on elemental features as input,[122] but these models tend to perform best when working with datasets much larger (e.g., > $10^5$ points) than those considered here (average dataset size = 2140 points). [122]

The results of the model fit accuracy assessment for all datasets are discussed in **Section 3.1**.

## 2.3. Assessment of model error bars

The model error bars used in this work are based on the work of Palmer et al.,[76] and are obtained using the ensemble of meta-estimators comprising the ML model. In the present work, our focus on using RF models means the predictions and error bars are obtained by taking the average and a calibrated version of the standard deviation of the predictions of the individual decision trees in the RF, respectively. The approach of Palmer et al. is general in the sense that it can be applied to any bagged regressor ML model, for example an ensemble of bagged linear regressors, bagged neural networks, etc. While one can obtain an error bar on each predicted data point by taking the standard deviation of the meta-estimators in the ensemble, one does not know whether these error bars are accurate. Ideally, the predicted errors should correlate perfectly with the true errors (i.e., residuals). The work of Palmer et al. developed an approach to assess the quality of the predicted error bars and used a linear scaling to recalibrate the error bars such that the distribution of z-scores (z-score = true error / predicted error) followed as closely as possible to a standard normal distribution. (More precisely, the negative log likelihood between a standard normal distribution and the distribution of z-scores is minimized.) The predicted error bars were shown to have equal or better correlation with the true errors after this recalibration for all cases tested. In this work, we use the approach of Palmer et al. implemented in MAST-ML together with additional types of analysis inspired by work from Pernot[123–125] and Schultz et al.[83] to assess the quality of the error bar recalibration for all 33 datasets. In practice, if one is satisfied with the error bar recalibration performance, only the recalibration parameters are needed for the final deployed model. If linear recalibration is performed, these parameters are a single slope and intercept which linearly transform the raw ensemble error estimates to their calibrated values. The error bar recalibration parameters are included in the Garden-hosted models, so any new predictions will automatically include calibrated error bars to quantify the uncertainty in the model prediction.

The results of the error bar assessment for all datasets are discussed in **Section 3.2**.

## 2.4. Assessment of model domain

In this work, we use the ML model domain of applicability approach recently proposed and assessed by Schultz et al.[83] Schultz et al. used a kernel density estimate (KDE) of the training data to determine when a predicted data point was too far from the training data and therefore out-of-domain. More specifically, they relate the distance of data points in feature space from the training data with two metrics of model quality: (i) model accuracy as determined by the reduced RMSE, which is the RMSE divided by the dataset standard deviation (RMSE/$\sigma_y$), and (ii) model precision as determination by miscalibration area ($E^{area}$), which is a measure of the accuracy of predicted error bars (determined by the Palmer method described in **Section 2.3**). Consider first RMSE/$\sigma_y$. Schultz et al. defined in- vs. out-of-domain by the values of RMSE/$\sigma_y$ being above/below a cutoff set by a simple naïve model, where worse (higher) RMSE/$\sigma_y$ values than a naïve model are taken to imply that the ML model is out-of-domain. They then showed that this categorization can be accurately predicted by the features for a data point being above/below a cutoff in KDE. The KDE fit to training data and this cutoff are established during model development and can then be used during inference to quantify any test data points as being in- vs. out-of-domain. This process can also be done for $E^{area}$, so this approach provides two separate (although correlated) determinations of in- vs. out-of-domain. The approach of Schultz et al. is integrated with MAST-ML, and in this work we determine KDE domain cutoffs for RMSE/$\sigma_y$ and $E^{area}$ for all 33 materials properties. The domain estimator is included in the Garden-hosted models, so any new predictions will automatically provide designation of whether the point should be considered as in-domain or out-of-domain based on each of these criteria.

The results of the model domain of applicability assessment for all datasets are discussed in **Section 3.3.**

### 2.5. Machine learning model hosting

In this work, we make use of the Garden-AI infrastructure to host our ML models in a way that is persistent, citable, accessible, and readily useable. Garden-AI containerizes the environment needed to run an ML model, including Python packages, system configuration, and model execution code. It links the containerized environment to serialized model files in a public model repository like Hugging Face or GitHub. Garden-AI then hosts the container on a public Docker registry. Once a model has been published with Garden-AI, other users can invoke the model remotely on a server that Garden-AI provides or on a local computer or a research computing cluster at their institution that has installed Globus Compute (previously funcX).[89] In this way, the user does not need to have specific knowledge of setting up Python environments, managing containers, using software repositories like Github, or supply local computing resources to run the model. Furthermore, models run regardless of changes to Python, packages, or operating systems that might make the original code unusable without significant changes. Overall, the Garden-AI infrastructure streamlines the use of numerous ML models in a single Python notebook, lowering the barrier for practical application of ML models. A Python notebook used to call each of the ML models hosted on Garden-AI is provided in the supporting information (see **Data and Code Availability**).

In **Section 3.4,** we demonstrate the use of several of our ML models to search for highly active and stable perovskite catalyst materials.

## 3. Results and Discussion

### 3.1. Machine learning model fit accuracy

In this section, we assess the predictive accuracy of our RF models for all 33 materials properties described in **Section 2.1**. **Figure 2A** contains example parity plots (plot of ML-predicted vs. true values) from 5-fold cross validation using MAST-ML for three different datasets. The 5-fold cross validation plots for all other datasets are available online (see **Data and Code Availability)**, and here we show three illustrative examples of datasets with excellent (phonon frequency, **Figure 2A**), modest (perovskite thermal expansion, **Figure 2B**), and somewhat poor fits (piezoelectric displacement, **Figure 2C**). For the phonon frequency dataset, the high $R^2$ and low mean absolute error (MAE) and RMSE indicate the RF model fits the dataset well. To help compare the quality of ML fits across many properties, we also report the reduced RMSE (RMSE/$\sigma_y$). An RMSE/$\sigma_y$ value of 1 (or greater) is indicative of approximately no predictive ability as the model is essentially guessing the mean value of the dataset. From **Figure 2**, the RMSE/$\sigma_y$ value of 0.297 for the phonon frequency dataset is significantly less than 1, indicating a statistically very good ML model fit. The progressively higher RMSE/$\sigma_y$ values of 0.553 and 0.827 for the perovskite thermal expansion and piezoelectric displacement datasets represent modest and somewhat poor-quality ML model fits, respectively. To assess the quality of all ML property models, in **Figure 3** we plot the RMSE/$\sigma_y$ from 5-fold cross validation for all 33 properties. We generally see robust ML fits with low errors for most properties, with 19/33 properties having extremely good fits of RMSE/$\sigma_y$ < 0.4, another 11/33 properties with modest fits of 0.4 < RMSE/$\sigma_y$ < 0.7, and only 3/33 properties showing poor fits with RMSE/$\sigma_y$ > 0.7. All key input and output files for all ML fits using MAST-ML are available online (see **Data and Code Availability**).

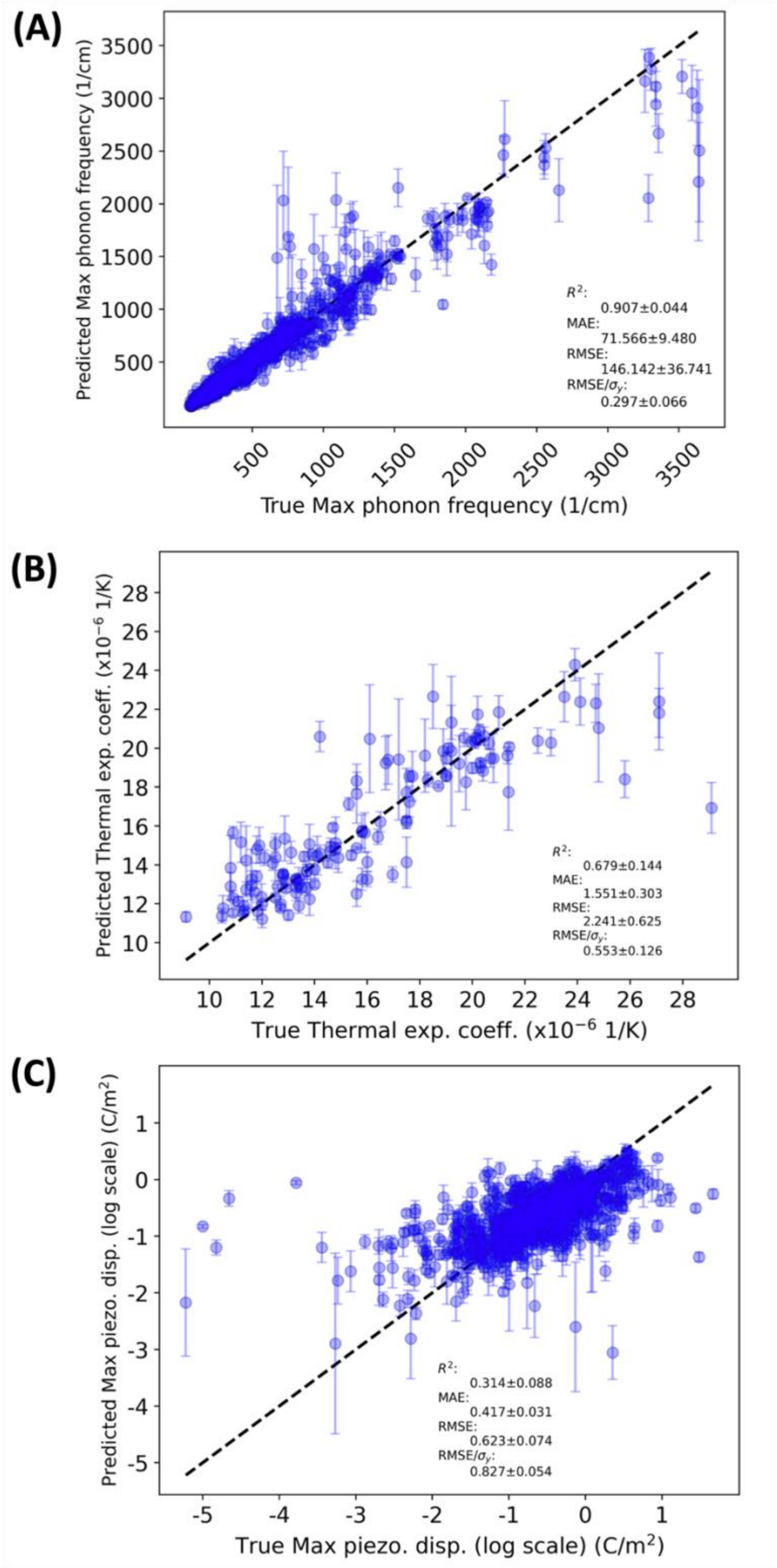

**Figure 2.** Example 5-fold cross validation parity plots. (A) parity plot for the phonon frequency dataset showing extremely good fit (RMSE/$\sigma_y$ < 0.4). (B) parity plot for the perovskite thermal

expansion dataset, showing modest fit (0.4 < RMSE/$\sigma_Y$ < 0.7). (C) parity plot for the piezoelectric dataset, showing somewhat poor fit (RMSE/$\sigma_Y$ > 0.7). The +/- metric values are the standard deviation across 25 splits of cross validation.

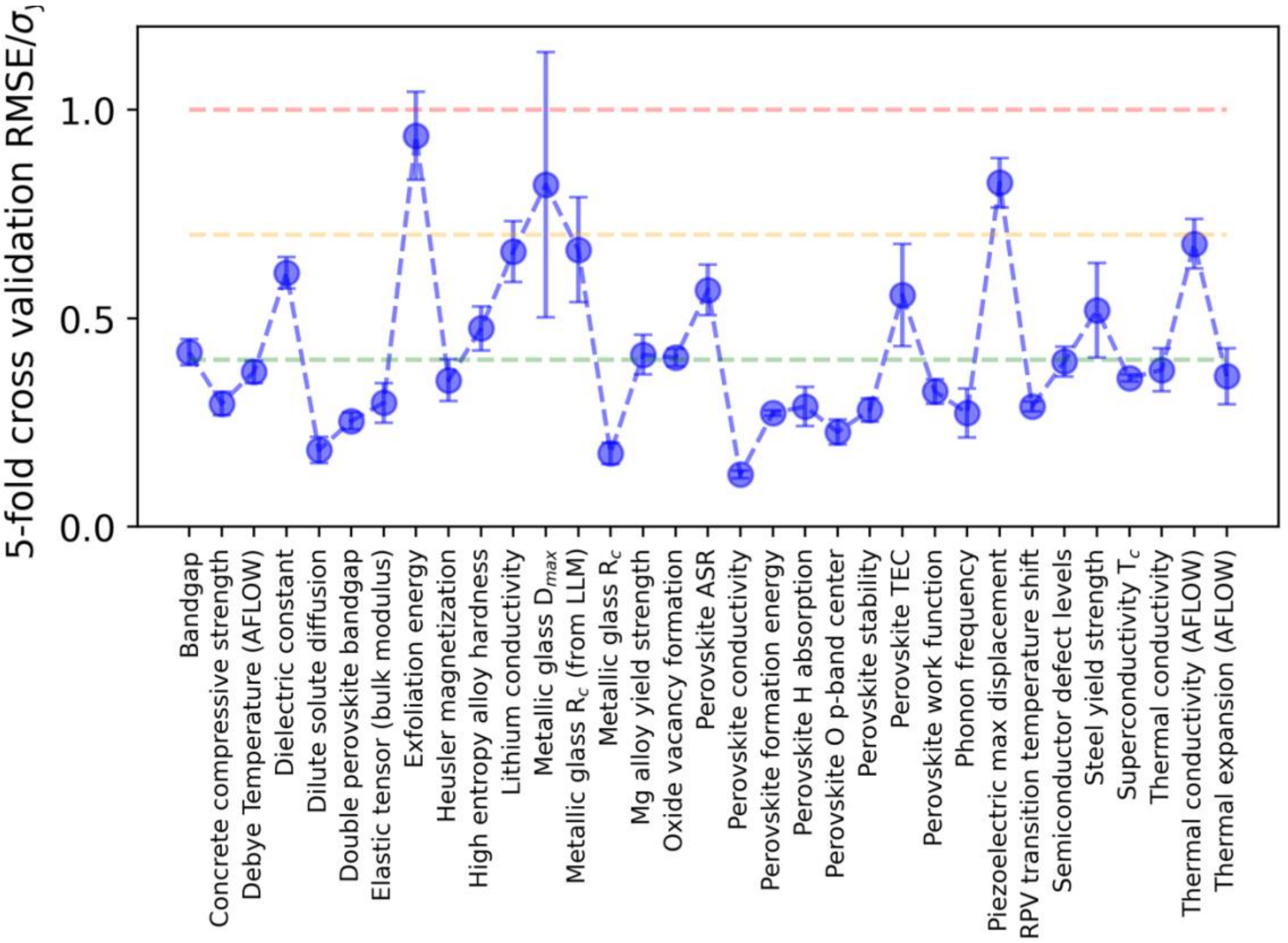

**Figure 3.** Summary of ML model performance represented as RMSE/$\sigma_Y$ from 5-fold cross validation for all 33 materials properties investigated in this work. The error bars are standard deviation of RMSE/$\sigma_Y$ across 25 splits of cross validation. The green, yellow, and red dashed lines are subjective thresholds for what RMSE/$\sigma_Y$ value constitutes a very good, modest, and poor model, respectively.

**Table 2** compares our RF model fits to other ML models fits for the same quantities and, where possible, the same datasets from the literature, including a detailed summary of the type and number of features used and key error metrics. There are cases where the literature reference did not fit to exactly the same data (e.g., an updated version may have been used in this work). Such cases are noted in **Table 2** because such a fit is still relevant but should not be quantitatively compared against the present results. In addition, direct comparison with literature values is complicated by using different models, feature sets, and assessment procedures (e.g., 5-fold vs. 10-fold cross validation, reporting of a single test split vs. average of many splits, etc.). Overall, out of the 33 properties fit in this work, 27 of them have literature reports to compare against. Of these 27, 16 of them were deemed to show good agreement with our present models, 5 of them show lower error than our models, so the reference model may be better (one of these, perovskite work functions, used DFT-based features to obtain lower error), 2 of them show higher error than our model, so our model may be better, and 4 of them had no direct comparison. While 5 of the 27 properties have lower errors for the reference model, the reductions in errors are typically modest (about 10-15%). We reiterate that our models have the advantage of having quantitative error bars and domain of applicability determination, the benefits of which may be worth roughly a 10-15% increase in prediction error for these particular properties, depending on the application.

**Table 2.** Summary of ML model fits to all materials property datasets and comparison to available references. The MAE and RMSE/$\sigma_y$ values (+/- values) are averages (standard deviations) across 25 splits of 5-fold cross validation.

| Property name | Feature type | Number of features | MAE (units) | RMSE/$\sigma_y$ | Reference ML performance | Reference | Notes |
|---|---|---|---|---|---|---|---|
| Bandgap | Elemental | 25 | 0.328 +/- 0.016 eV | 0.419 +/- 0.031 | MAE of 0.447 eV (RF) | 53 | Our model has lower error |
| Concrete compressive strength | Material-specific | 8 | 3.413 +/- 0.241 MPa | 0.298 +/- 0.030 | RMSE of about 5 MPa (RF), agrees with our RMSE of 4.945 +/- 0.423 MPa | 91 | Agrees |
| Debye Temperature (AFLOW) | Elemental | 25 | 42.353 +/- 1.571 K | 0.371 +/- 0.027 | MAE of 36.48 K (RF) | 53 | Reference model has lower error |
| Dielectric constant | Elemental | 25 | 0.113 +/- 0.008 (log scale) | 0.610 +/- 0.038 | MAE of 0.271 (fit to refractive index, MODNet) | 59 | Cannot directly compare because models fit to different target data |
| Dilute solute diffusion | Elemental | 25 | 0.177 +/- 0.019 eV | 0.184 +/- 0.031 | RMSE of 0.129 eV (GPR), fit to shifted data | 126 | Cannot directly compare because models fit to different target data |
| Double perovskite bandgap | Elemental | 25 | 0.278 +/- 0.020 eV | 0.254 +/- 0.021 | RMSE of 0.37 eV (KRR), our RMSE is 0.399 +/- 0.026 eV | 95 | Agrees |
| Elastic tensor (bulk modulus) | Elemental | 25 | 12.511 +/- 1.075 GPa | 0.298 +/- 0.048 | MAE of about 1.12 GPa (coNGN), used structure data and about 10× more data | 59 | Cannot directly compare because models fit to different target data |
| Exfoliation energy | Elemental | 25 | 52.082 +/- 8.855 eV/atom | 0.938 +/- 0.106 | MAE of 55.105 eV/atom (RF) | 53 | Agrees |
| Heusler magnetization | Elemental + one-hot encoding of Heusler structure type | 25 | 95.938 +/- 10.033 emu/cm$^3$ | 0.476 +/- 0.053 | n/a | n/a | No reference model |
| High entropy alloy hardness | Elemental | 25 | 56.163 +/- 7.704 (HV) | 0.351 +/- 0.050 | RMSE of 75 MPa (SVR), agrees with our RMSE of 79.703 +/- 12.599 (HV) | 65 | Agrees |
| Lithium conductivity | Elemental | 25 | 0.902 +/- 0.119 S/cm (log scale) | 0.660 +/- 0.073 | MAE of 1.10 (autosklearn), MAE of 0.85 (CrabNet) S/cm (log scale) | 98 | Agrees |
| Metallic glass $D_{max}$ | Elemental | 25 | 2.364 +/- 0.286 (mm) | 0.820 +/- 0.310 | RMSE of 3.3 +/- 0.1 (GBR), used additional physical features, our RMSE is 5.395 +/- 1.699 (mm) | 127 | Agrees, though our RMSE has much larger range |
| Metallic glass $R_c$ (from LLM) | Elemental | 25 | 0.680 +/- 0.141 K/s (log scale) | 0.665 +/- 0.126 | n/a | n/a | No reference model |
| Metallic glass $R_c$ | Elemental + one-hot encoding of data type | 25 | 0.129 +/- 0.017 K/s (log scale) | 0.176 +/- 0.026 | RMSE of 0.36 K/s (log scale) (RF), our RMSE is 0.389 +/- 0.055 | 101 | Agrees |
| Mg alloy yield strength | Material-specific | 14 | 26.929 +/- 3.259 MPa | 0.382 +/- 0.051 | RMSE of 35 +/- 5 MPa (RF), our RMSE is 41.115 +/- 5.020, reference did 10-fold CV | 102 | Agrees |
| Oxide vacancy formation | Elemental | 25 | 1.081 +/- 0.037 eV | 0.405 +/- 0.021 | n/a | n/a | No reference model |
| Perovskite ASR | Elemental + one-hot encoding of electrolyte type + ML-predicted barrier | 25 | 0.466 +/- 0.052 Ohm-cm$^2$ (log scale) | 0.568 +/- 0.061 | RMSE of 0.590 +/- 0.05, our RMSE is 0.606 +/- 0.077 Ohm-cm$^2$ (log scale) | 57 | Agrees |
| Perovskite conductivity | Elemental + material-specific | 25 | 0.139 +/- 0.005 S/cm (log scale) | 0.126 +/- 0.009 | RMSE of 0.24 S/cm (XGBoost, log scale, 10% leave-out), our RMSE is 0.264 +/- 0.018 S/cm (log scale) | 105 | Agrees |
| Perovskite formation energy | Elemental | 25 | 0.114 +/- 0.003 eV/atom | 0.273 +/- 0.007 | MAE of 0.152 eV/atom (RF) | 53 | Our model has lower error |
| Perovskite H absorption | Elemental + material-specific | 25 | 0.010 +/- 0.001 mol/(formula unit) | 0.288 +/- 0.047 | RMSE of 0.021 (XGBoost), our RMSE is 0.020 +/- 0.004 (mol/(formula unit)) | 107 | Agrees |
| Perovskite O p-band center | Elemental | 50 | 0.146 +/- 0.009 eV | 0.227 +/- 0.030 | n/a | n/a | No reference model |
| Perovskite stability | Elemental | 50 | 29.220 +/- 1.477 meV/atom | 0.280 +/- 0.028 | RMSE of 29 meV/atom (KRR), our RMSE is 53.436 +/- 5.148 meV/atom | 128 | Cannot directly compare as previous fit had ~1000 fewer points and was under different thermodynamic conditions |
| Perovskite TEC | Elemental | 25 | 1.469 +/- 0.331 ($\times 10^{-6}$ $K^{-1}$) | 0.556 +/- 0.123 | RMSE of 1.54 ($\times 10^{-6}$ $K^{-1}$) (SVM, 10-fold CV), our RMSE from 10-fold CV is 2.088 +/- 0.680 ($\times 10^{-6}$ $K^{-1}$) | 109 | Agrees, though our per-split RMSE has large range due to small dataset size. |
| Perovskite work function | Elemental | 25 | 0.430 +/- 0.036 eV | 0.324 +/- 0.029 | RMSE of 0.468 eV (RF), our RMSE is 0.562 +/- 0.050 eV | 110 | Reference model has lower error, but uses DFT-based features that are difficult to calculate |
| Phonon frequency | Elemental | 25 | 62.928 +/- 6.492 $cm^{-1}$ | 0.273 +/- 0.059 | MAE of 68.687 $cm^{-1}$ (RF) | 53 | Agrees |
| Piezoelectric max displacement | Elemental | 25 | 0.411 +/- 0.030 $C/m^2$ (log scale) | 0.825 +/- 0.060 | n/a | n/a | No reference model |
| RPV transition temperature shift | Material-specific + one-hot encoding of product form | 15 | 9.170 +/- 0.284 °C | 0.289 +/- 0.012 | RMSE of 12.2 +/-- 0.7 °C (NN ensemble), our RMSE is 13.300 +/- 0.397 °C | 113 | Agrees |
| Semiconductor defect levels | Elemental + material-specific | 25 | 0.361 +/- 0.029 eV | 0.396 +/- 0.036 | RMSE of 0.42 eV (GBR, 10-fold CV), our RMSE is 0.509 +/- 0.041 eV | 114 | Reference model has lower error |
| Steel yield strength | Elemental | 25 | 99.178 +/- 8.444 MPa | 0.519 +/- 0.114 | MAE of 82.3 MPa (AutoML-Mat), RF-Magpie and CrabNet have MAE of 103.5 and 107.3 MPa, respectively | 59 | Agrees when comparing similar models. Best reference model has lower error |
| Superconductivity $T_c$ | Elemental | 50 | 0.166 +/- 0.004 K (natural log scale) | 0.356 +/- 0.008 | $R^2$ of 0.88 (RF, 15% held out data), our $R^2$ is 0.873 +/- 0.006 | 116 | Agrees |
| Thermal conductivity | Elemental + material-specific | 25 | 0.109 +/- 0.012 W/m-K (log scale) | 0.376 +/- 0.052 | n/a | n/a | No reference model |
| Thermal conductivity (AFLOW) | Elemental | 25 | 2.962 +/- 0.141 W/m-K | 0.679 +/- 0.059 | MAE of 2.658 W/m-K (RF) | 53 | Reference model has lower error |
| Thermal expansion (AFLOW) | Elemental | 25 | $5.9\times10^{-6}$ $K^{-1}$ | 0.361 +/- 0.067 | MAE of $5.44\times10^{-6}$ $K^{-1}$ (RF) | 53 | Reference model has lower error |

### 3.2. Machine learning model error bars

In this section, we discuss how one may use the method of Palmer et al., discussed in **Section 2.3** and integrated into our MAST-ML code, to assess the quality of model error bars. We then show how our approach for generating well-calibrated error bars performs on all 33 materials properties. **Figure 4** shows four plots which can be used to assess the quality of the error bars prior to, and after, calibration, using the concrete compressive strength dataset as an illustration. Equivalent plots for all other materials properties are available online (see **Data and Code Availability**).

**Figure 4A-C** focuses on evaluating the distribution of z-score values and the effectiveness of calibrating the error bars to obtain an improved z-score distribution (i.e., closer to standard normal distribution). **Figure 4A** shows the z-score distribution for all left-out test data. Prior to calibration, the mean and standard deviation of the z-scores are 0.025 and 0.871, respectively. After calibration, the values shift to 0.017 and 0.966, respectively. Ideally, the z-score distribution is standard normal, meaning it has a mean of 0 and standard deviation of 1. For this case, recalibration of the error bars resulted in an improved distribution of the z-scores. For the remaining plots, the data are binned into tranches of values based on their reduced model error bar values, i.e., the model error bars divided by the standard deviation of the dataset. **Figure 4B**

provides another look at the z-score distributions, but this time the z-scores are plotted as a function of the binned reduced model error values, with a constant number of points per bin. Such a visualization can provide additional information than what is depicted in **Figure 4A**, as it shows the mean and standard deviation of the z-scores on a per-bin basis, enabling one to see if the z-score distribution deviates from ideal value at a particular error bar magnitude. In **Figure 4B**, the average per-bin z-score and average per-bin standard deviation of the z-score have values of 0.0264 and 0.8037 prior to calibration and 0.0207 and 0.8946 after calibration, again demonstrating that calibration improves the distribution of z-scores. **Figure 4C** is a representation of the "error in the predicted error" before and after calibration, quantified as the magnitude of the area between cumulative distribution function (CDF) curves of the z-score distribution and a standard normal distribution, where ideally a miscalibration area of 0 corresponds to a perfect, infinitely sampled, standard normal z-score distribution. The distance of the points in **Figure 4C** from 0 directly correspond to how far away the average and standard deviation of the points in **Figure 4B** are from 0 and 1, respectively. Overall, in **Figure 4C** we again see that calibration produces more accurate error bars, where before and after calibration, the average per-bin miscalibration area is 0.049 and 0.045, respectively.

**Figure 4D** shows an evaluation of the quality of the correlation between true and predicted errors and the effectiveness of calibrating the error bars to improve this correlation. **Figure 4D** is a plot of the reduced root mean square residuals vs. binned reduced model error estimates (in our prior studies we refer to this as a "residual vs. error" or "RvE" plot).[1,57,76,113] This plot is a direct assessment of how well, on average, the true error and predicted error correlate with each other. Ideally, the predicted error would be perfectly correlated with the true error, meaning the slope and intercept in **Figure 4D** would be equal to 1 and 0, respectively. From **Figure 4D**, prior to calibration, the slope and intercept are 0.65 and 0.06, respectively, indicating that the uncalibrated error bars tend to be overestimating the true error. After calibration, the slope and intercept shift to 1.02 and -0.01, respectively, indicating a large improvement in the correlation of true vs. predicted errors after calibration.

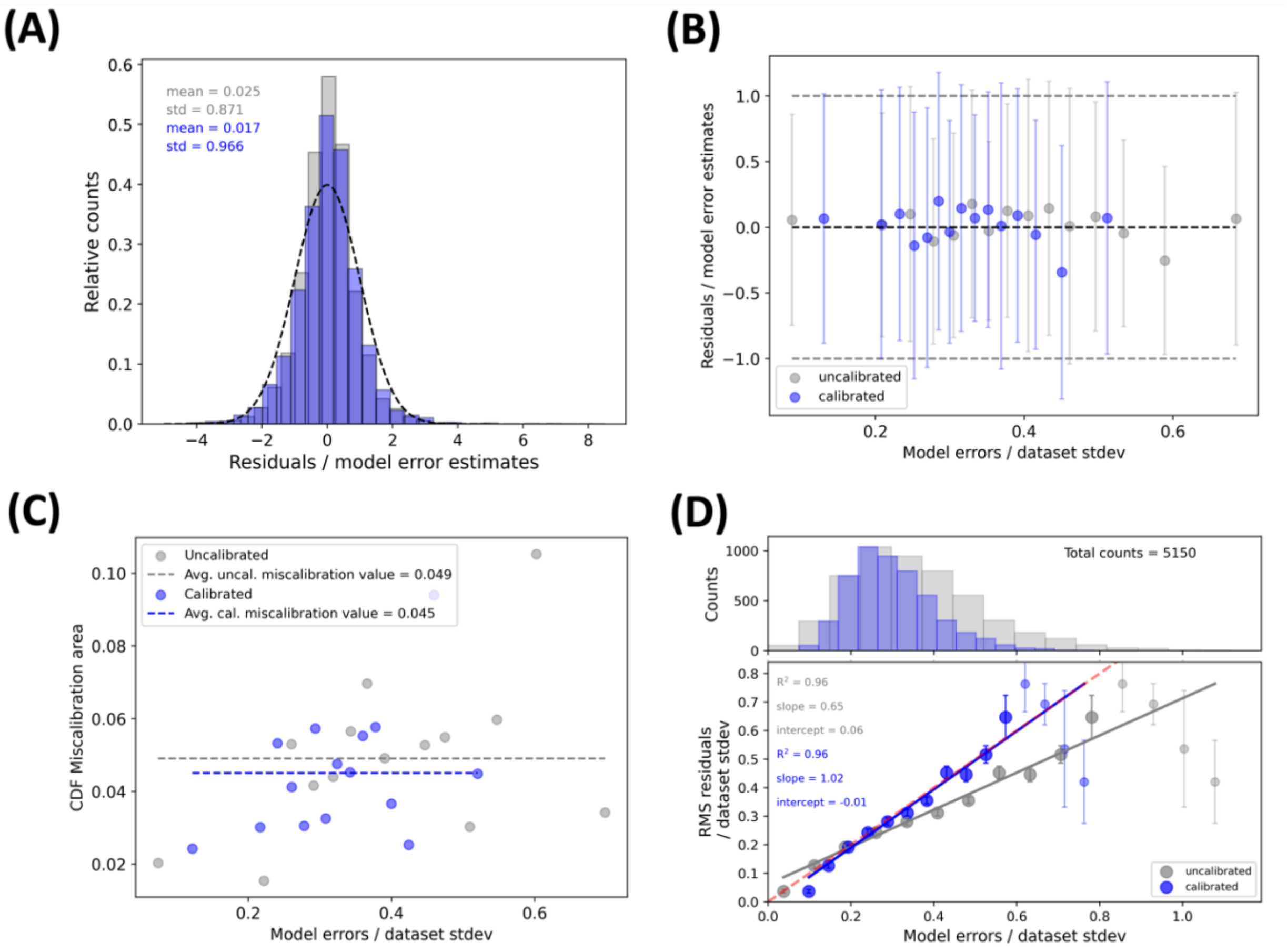

**Figure 4.** Plots used to assess the quality of error bars before and after calibration. For all plots, the grey and blue data correspond to data before and after calibration, respectively. These plots are for the concrete compressive strength dataset. (A) Distribution of z-scores over all test data. (B) z-score values binned by the reduced model error values. The error bars are the standard deviation of the z-scores for each bin. (C) The miscalibration area from the cumulative distribution function (CDF) of z-scores vs. binned reduced model error values. (D) Reduced root mean square residuals vs. binned reduced model error estimates. The histogram at the top of the plot shows the distribution of points present in each bin. The smaller, more transparent points correspond to bins with inadequate data for accurate sampling and therefore are not expected to yield accurate RMS residuals. In (B) and (C), the plots are made with a constant number of points per bin.

**Figure 5A** shows the behavior of the overall z-score distributions before (grey points and bars) and after (blue points and bars) calibration across all 33 datasets. The grey and blue points strongly overlap in all cases and hover near a mean z-score of 0. It is more obvious to see the improvement in the standard deviation of the z-score distributions after calibration, where the per-dataset z-score standard deviation after calibration is close to 1 for almost all cases. Through the lens of z-score distributions, calibration of the error bars resulted in improvement of the error bars for 28 of the 33 datasets, as determined by assessing the z-score standard deviation. In particular, the experimental bandgap, elastic tensor, exfoliation energy, and all three metallic glass datasets showed rather large z-score standard deviations that were dramatically corrected with calibration. **Figure 5B** shows the impact of calibration regarding the correlation of the true and predicted errors based on the slope and intercept from plots of the reduced RMS residuals vs. reduced model error estimates (e.g., as shown in **Figure 4D**). Perfect error bar correlation would result in slopes equal to 1 and intercepts equal to 0. Prior to calibration, most datasets show slopes significantly less than 1 and intercepts somewhat greater than 0. Across all datasets, the average slope and intercept before calibration is 0.76 and 0.09, respectively. After calibration, many datasets see their true vs. predicted error correlation slope move closer to 1, and intercept move closer to 0, where the average slope and intercept after calibration is 0.94 and 0.01, respectively. Through the lens of true vs. predicted error correlations, the calibration of error bars improved the correlation for 30 of the 33 datasets, as determined by the slope of the true vs. predicted errors. Overall, calibration improved the z-score distribution or the true vs. predicted error correlation for all datasets, demonstrating the effectiveness of the error bar calibration approach.

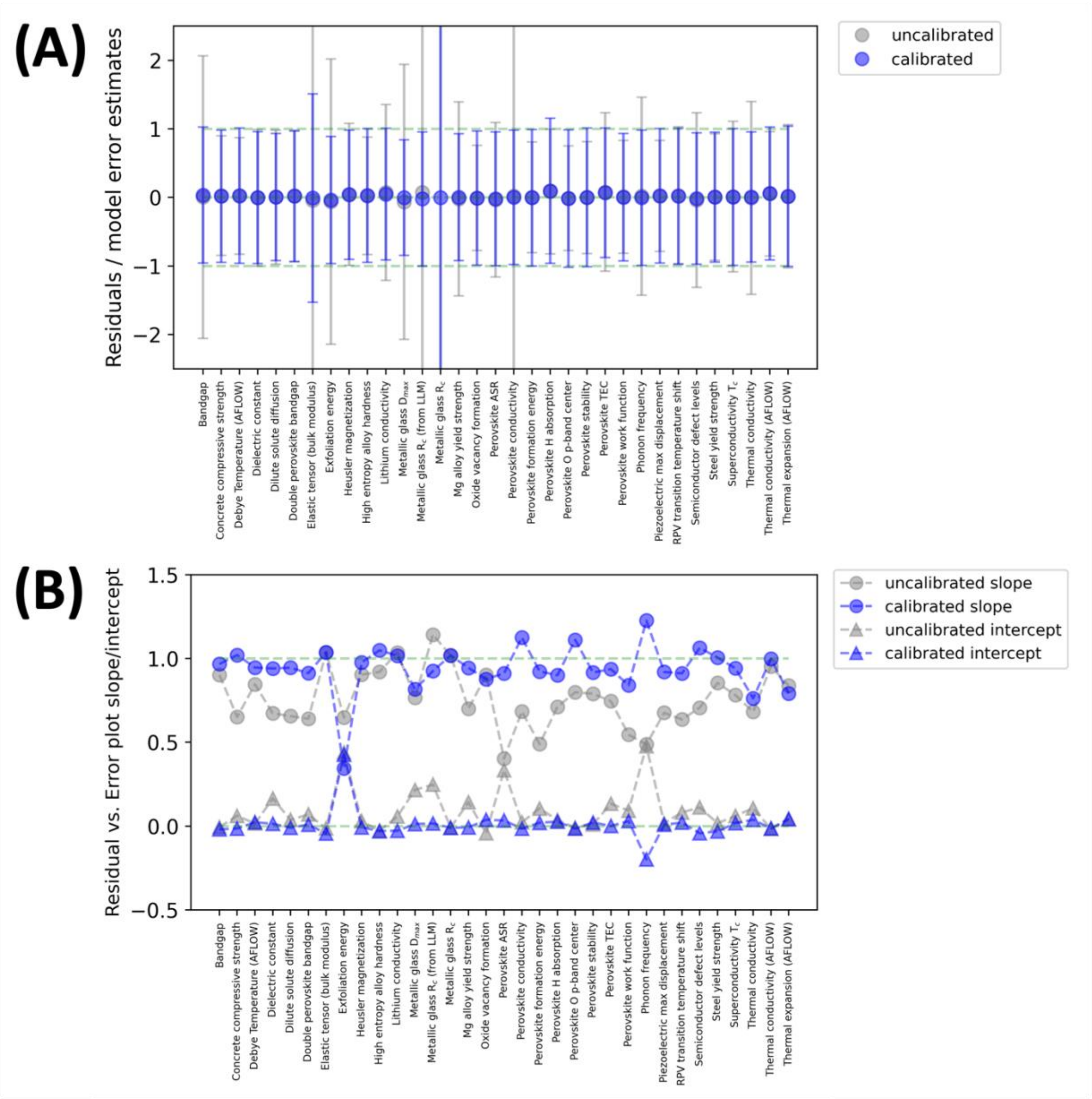

**Figure 5.** Summary of key error bar metrics before and after calibration for all 33 materials properties investigated in this work. For both plots, the grey and blue data correspond to data before and after calibration, respectively. (A) Assessment of z-score distributions. The points (error bars) represent average (standard deviation) of the overall z-score distributions. The dashed green lines are positioned at values of -1, 0 and 1 as a guide to the eye. (B) Assessment of true vs. predicted error correlation. The circle (triangle) points denote the slope (intercept) values from the correlation of reduced RMS residuals vs. reduced model error estimates (e.g., as shown in **Figure 4D**). The dashed green lines are positioned at values of 1 and 0 as a guide to the eye.

### 3.3. Machine learning model domain

We now discuss how one may use the method of Schultz et al., discussed in **Section 2.4** and integrated into our MAST-ML code, to assess ML model domain of applicability. **Figure 6** shows five plots that can be used to assess the quality of the domain-of-applicability approach based on thresholds of reduced RMSE (for prediction accuracy) and error bar miscalibration area (for error bar accuracy). These plots correspond to the lithium conductivity dataset. Equivalent plots for all other materials properties are available online (see **Data and Code Availability**).

**Figure 6A** and **Figure 6B** contain plots of binned reduced RMSE and binned miscalibration area as a function of KDE distance D from the training data, respectively. Each of the 10 bins in each plot contains approximately the same number of points. These plots both show qualitatively that test data sampled further from the training data distribution (i.e., larger D value) results in a model with less accuracy (i.e., higher reduced RMSE) or less robust error quantification (i.e., higher miscalibration area), indicating that, at sufficiently high D, the model may be unable to provide accurate predictions or accurate error bars for the given test data. For the assessment of reduced RMSE (miscalibration area), a D value of 0.99 (0.88) results in the best classification performance of data that is in- vs. out-of-domain using the thresholds denoted with the solid green lines in **Figure 6**. **Figure 6C** contains a confusion matrix showing the ability of our domain approach to successfully classify data points that are in- vs. out-of-domain. This confusion matrix is constructed by determining D thresholds like those chosen from **Figure 6A** and **Figure 6B** for subsets of data generated from 5-fold cross validation and running the same domain workflow as in **Section 2.4**. This is a check on the sensitivity of selecting D. For the reduced RMSE test given the lithium conductivity dataset shown here, the model can correctly classify domain about 90% of the time, and with only a 0.04% false positive rate of flagging an out-of-domain data point as erroneously in-domain. The analogous confusion matrix for the miscalibration area test is shown in **Figure 6D,** where the model can correctly classify domain about 96% of the time, and with only a 2% false positive rate of flagging an out-of-domain data point as erroneously in-domain. **Figure 6E** shows the correlation of reduced RMSE and miscalibration area with KDE distance D, showing that, in general, higher values of reduced RMSE (i.e., less accurate fit) and higher values of miscalibration area (i.e., less accurate error bars) correspond to higher D value and thus points which are further from the model domain of applicability.

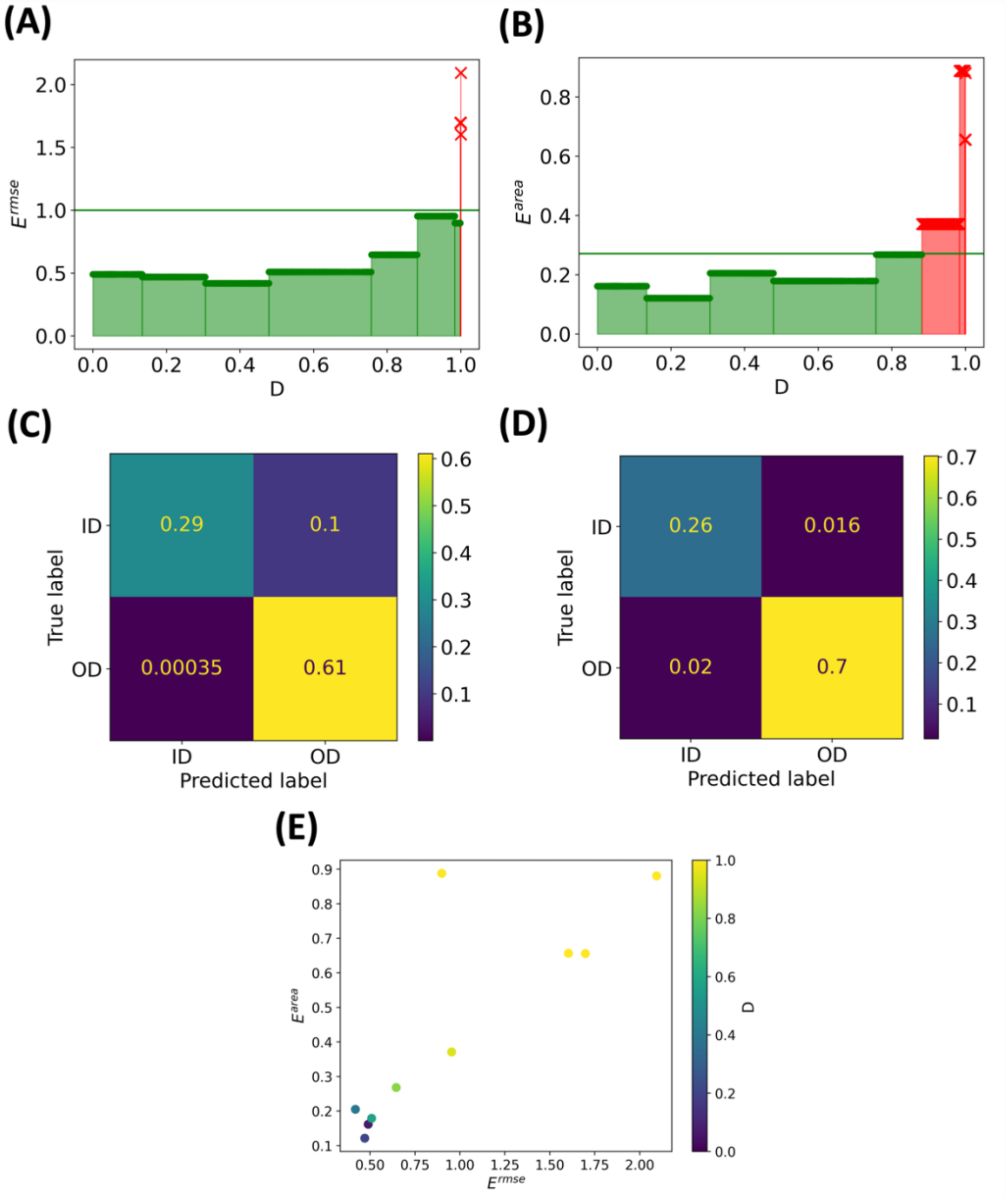

**Figure 6.** Five plots used to assess model domain of applicability, here for the lithium conductivity dataset. (A) Plot of binned reduced RMSE vs. KDE feature distance. (B) Plot of binned reduced error bar miscalibration areas vs. KDE feature distance. In (A) and (B), the green points denote in domain and the red point denote out of domain. The green line is the ground truth threshold between in and out of domain. (C) Confusion matrix for classification of in- vs. out-of-domain, expressed as a fraction of the total data. (D) Confusion matrix for the classification of in- vs. out-of-domain, expressed as a fraction of the total data. (E) Plot of reduced error bar miscalibration area vs. reduced RMSE with color denoting KDE feature distance.

**Figure 7** provides an overall assessment of our domain approach to classify test data correctly for all materials properties. Here, by "classify correctly", we mean the ability of the domain model to flag correctly in-domain and out-of-domain points as being in-domain and out-of-domain, respectively. Inspecting the confusion matrices for the miscalibration area and reduced RMSE domain classification criteria for all materials properties reveals that 25/33 and 30/33 correctly classify data with >80% accuracy, respectively. These numbers drop to 12/33 and 20/33 when considering >90% classification accuracy. Overall, we find the approach of Schultz et al. broadly provides informative measures of model domain of applicability based on both ML model predictions (reduced RMSE criterion) and ML model error bar accuracy (miscalibration area criterion) for almost all materials properties.

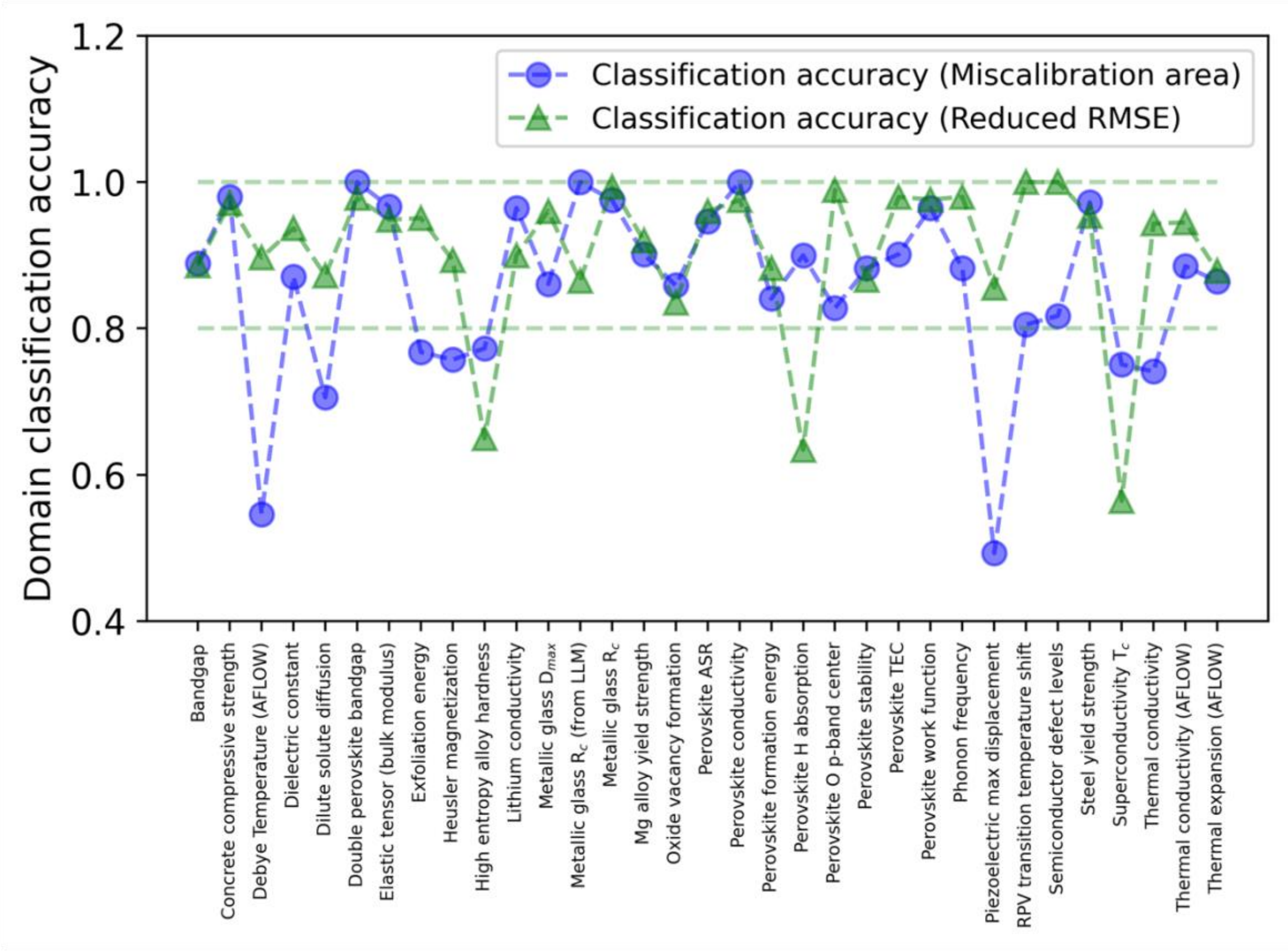

**Figure 7.** Summary of domain classification accuracy for all 33 datasets examined in this work based on the miscalibration area (blue circles) and reduced RMSE (green triangles) criteria. The horizontal dashed lines at a classification accuracy of 1.0 and 0.8 are guides for the eye to denote perfect and very good domain classification accuracies, respectively.

### 3.4. Materials discovery example using Garden-hosted models

The availability of high-quality ML property models with associated uncertainties and domain guidance can enable new approaches to materials screening. To illustrate how these models can be applied in practice, we report here on a fully ML-based materials screening exercise in which we evaluate a large number of perovskite oxide materials in search of stable and highly active oxide catalysts with good compatibility with known electrolytes. All screened datasets obtained from this analysis are available online (see **Data and Code Availability**).

This analysis extends previous work in which we screened more than 19 million perovskite oxide compositions based on cost, perovskite thermodynamic stability, and perovskite ASR values.[57] Here, we extend that work to include additional materials properties relevant for oxide catalysts: perovskite formation energy (used to complement the perovskite thermodynamic stability model), perovskite thermal expansion coefficient (used to inform electrode compatibility with commonly-used electrolytes), and perovskite conductivity (used with the perovskite ASR model to further inform expected catalytic activity). In addition, we use the available domain of applicability guidance for each model to help restrict our search space to those materials deemed inside the ML model domain.

We first use our ML models to predict properties for six baseline perovskite catalyst materials, with values shown in **Table 3**. The chosen baseline materials are $La_{0.6}Sr_{0.4}CoO_3$ (LSC), $La_{0.6}Sr_{0.4}Co_{0.2}Fe_{0.8}O_3$ (LSCF), $Ba_{0.5}Sr_{0.5}Co_{0.8}Fe_{0.2}O_3$ (BSCF), $Sr_{0.9}Cs_{0.1}Co_{0.9}Nb_{0.1}O_3$ (SCCN), $BaFe_{0.125}Co_{0.125}Zr_{0.75}O_3$ (BFCZ), and $BaFe_{0.2}Co_{0.2}Zr_{0.1}Y_{0.1}O_3$ (BFCZY). These materials were chosen as they represent logical benchmarks for which to conduct materials screening, as discussed more below. Each of these materials is well-studied, so qualitative trends of many of their properties are known (e.g., trend of ASR of LSCF vs. BSCF vs. BFCZY). LSCF is a commercial cathode material in SOFCs; BSCF is a non-commercial but often-used benchmark high-performing material; and BFCZ, BFCZY, and SCCN are all more recently proposed high-performing materials for SOFCs/SOECs and PCFCs.[55,129,130] The ML stability, conductivity, TEC, and ASR predictions for these baseline materials agree with established knowledge in the field. Regarding stability, BSCF is less stable than LSCF, and BFCZY (with combined 20% of Zr and Y) is less stable than BFCZ, which has 75% Zr.[130,131] There is no clear trend of the predicted formation energies and the stabilities, which may be because the database used to train the formation energy model was comprised of mostly oxyhalide perovskites. Regarding conductivity, LSC, LSCF and BSCF are well-known to be good mixed ionic-electronic conductors with high conductivities, and these materials have predicted log conductivities in the range of about 1.5-2.5 S/cm at 500 °C.[132–134] On the other hand, the large amount of Zr present in BFCZ significantly reduces its conductivity by at least two orders of magnitude compared to LSCF, a result in agreement with recent studies investigating this material as both a stand-alone electrode and as a composite electrode for fuel cells.[131,135]

In the case of TEC, it is known that materials with high amounts of Co have higher TECs of about $20\times10^{-6}$ $K^{-1}$, compared to those with reduced amounts of Co.[109,132,133] We observe that LSC, BSCF and SCCN, all of which have at least 80% Co on the B-site, have higher TECs than materials like LSCF and BFCZ, which have less Co.[132–134] The choice of electrolyte will be influenced not only by the thermodynamic stability with the electrode material but also by the difference in TEC between electrolyte and electrode, as minimizing TEC mismatch is advantageous to deter delamination and improve overall performance. Regarding predicted ASR, it is known that the materials BSCF, SCCN, BFCZ, and BFCZY are all more active than LSC and LSCF, which is reflected in the much smaller predicted ASR values for these materials compared to LSC and LSCF. It is interesting to note that BFCZ has an ASR that is comparable to those for BSCF and BFCZY despite having a high Zr content of 75%, in qualitative agreement with findings from previous work.[131]

**Table 3.** Summary of ML-predicted property values for baseline well-studied perovskite catalyst materials.

| **Material** | **Stability (meV/atom)** | **Formation energy (eV/atom)** | **TEC ($\times10^{-6}$ 1/K)** | **Conductivity (log scale, T=500 °C) (S/cm)** | **ASR (log scale, ceria electrolyte, T=500°C) (Ohm-cm²)** |
|---|---|---|---|---|---|
| LSC | 60.6 | 0.56 | 19.1 | 2.5 | 1.44 |
| LSCF | 44.0 | 0.53 | 16.8 | 1.9 | 1.33 |
| BSCF | 126.7 | 0.36 | 20.0 | 1.7 | 0.21 |
| SCCN | 130.8 | 0.32 | 21.5 | 1.9 | 0.01 |
| BFCZ | 18.5 | 0.42 | 17.1 | -1.3 | 0.49 |
| BFCZY | 62.3 | 0.53 | 19.7 | 0.4 | 0.19 |

Next, we use our ML models to predict stability for 19,072,821 candidate perovskite compositions. More information on how this list of candidate materials was chosen is available from Ref. 57. **Figure 8A** contains a violin plot showing the distribution of predicted stability values at various stages of screening, where the stability values are predicted at 500 °C, a benchmark low temperature for design of fuel cells. The leftmost violin shows the full distribution of predicted stability values, which range from a minimum of 0.09 meV/atom to a maximum of 1320.5 meV/atom for the full set of materials. The second violin screens out materials with stability values deemed out-of-domain based on the reduced RMSE criterion (i.e., their predictions are seen as not reliable), leaving 9,701,688 materials (50.9 %) as promising. The third violin further screens the stability distribution by removing those values which are above 100 meV/atom, an approximate threshold of instability, leaving just 836,386 materials (4.4%) as promising. It is worth noting that if the domain guidance is not invoked, then 1,353,383 materials (7.1%) remain after removing those which are above 100 meV/atom, indicating there are 516,997 materials which were predicted to be stable yet marked as out-of-domain. It is striking that simply invoking the domain-of-applicability model for stability predictions removed half of our prospective materials from contention, while selection of a sensible stability cutoff resulted in further drastic reduction of prospective materials to only about 4% of our initial pool.

Next, we focus on the 836,386 predicted stable, in-domain materials and use our ML model of ASR to predict the ASR (assuming a ceria electrolyte) at 500 °C for these materials, as shown in **Figure 8B.** The leftmost violin shows the full distribution of predicted log ASR values, which range from a minimum -0.47 Ohm-cm$^2$ to a maximum of 3.89 Ohm-cm$^2$. Similarly to the stability analysis, the second violin screens out materials with ASR values deemed out-of-domain based on the reduced RMSE criterion, leaving 505,735 (2.7%) as promising. The third violin further screens the ASR distribution by removing those values which are above 0.2 Ohm-cm$^2$ (the value of high-performing material BSCF from **Table 3**), leaving just 12,530 materials (0.07%) as promising.

It is interesting to analyze further this set of 12,530 predicted stable, highly active materials based on their predicted conductivity (**Figure 8C**) and TEC (**Figure 8D**) values. The conductivity values range over orders of magnitude, from a maximum log conductivity of 2.7 S/cm and a minimum of -4.5 S/cm. Materials with high conductivity would be suitable as standalone, single-phase electrodes while those with low conductivity would likely need to be incorporated into a composite electrode to enable sufficient conductivity to realize the low predicted ASR. When examining the subset of materials with low log conductivities less than -1.3 S/cm (the value of BFCZ from **Table 3**), we find there are 129 such materials, all of which have predicted log ASR values less than 0.5 Ohm-cm$^2$, the value for BFCZ. Given that BFCZ with 75% Zr has been successfully integrated with LSCF to make a well-performing composite electrode,[131] the present screening analysis suggests many more materials, when incorporated as composite electrodes, may enable even better performance. Some top candidate materials suitable for making composite electrodes are $BaNb_{0.125}Co_{0.25}Sn_{0.375}Mo_{0.25}O_3$, $BaZr_{0.125}Nb_{0.375}Co_{0.25}Sn_{0.25}O_3$, and $Ca_{0.25}Bi_{0.125}K_{0.625}Co_{0.5}Sc_{0.125}Fe_{0.375}O_3$, with predicted log ASR values of -0.08, -0.01 and -0.01 Ohm-cm$^2$, respectively. On the other hand, there are 1346 materials with high log conductivities, greater than 2 S/cm, which are promising single phase electrode materials. Amongst those materials, top candidates that also have low ASR values are $Sr_{0.75}Ba_{0.125}Sm_{0.125}Co_{0.75}Sc_{0.125}Ni_{0.125}O_3$ (predicted log ASR = -0.3 Ohm-cm$^2$) and $K_{0.125}Sm_{0.125}Sr_{0.75}Y_{0.125}Ni_{0.125}Co_{0.75}O_3$ (predicted log ASR = -0.2 Ohm-cm$^2$).

Finally, let us consider the TEC values shown in **Figure 8D**. The TEC values for the 12,530 stable, low-ASR materials range from a minimum of $14.6\times10^{-6}$ K$^{-1}$ to a maximum of $23.8\times10^{-6}$ K$^{-1}$. We note that the TEC values for the most commonly used electrolytes are roughly in the range of 9-13$\times10^{-6}$ K$^{-1}$ (depending on electrolyte composition and temperature): yttria-stabilized zirconia (YSZ) has a value of $9\times10^{-6}$ K$^{-1}$, [136] perovskite $La_{0.9}Sr_{0.1}Ga_{0.8}Mg_{0.2}O_{3-\delta}$ (LSGM) has a value of 9-10$\times10^{-6}$ K$^{-1}$,[137] and gadolinium-doped ceria (GDC) has a value of 11-12$\times10^{-6}$ K$^{-1}$.[138] Examination of the domain of applicability yields only 177 of the 12,530 materials as in-domain for TEC based on the reduced RMSE criterion. The small number of in-domain materials is likely a reflection of the small training dataset of only 137 data points for the TEC model. Focusing just on this small set of 177 materials, only 14 have predicted TEC values less than $17\times10^{-6}$ K$^{-1}$. The material with the smallest predicted TEC ($14.6\times10^{-6}$ K$^{-1}$) is $Sr_{0.5}Bi_{0.125}Pr_{0.375}Y_{0.125}Ni_{0.125}Fe_{0.75}O_3$. Its predicted log ASR of 0.2 Ohm-cm$^2$ and predicted log conductivity of 1.2 S/cm make it an interesting predominantly Fe-based electrode material that may be expected to have activity comparable to BSCF but with improved thermodynamic and electrolyte interface stability.

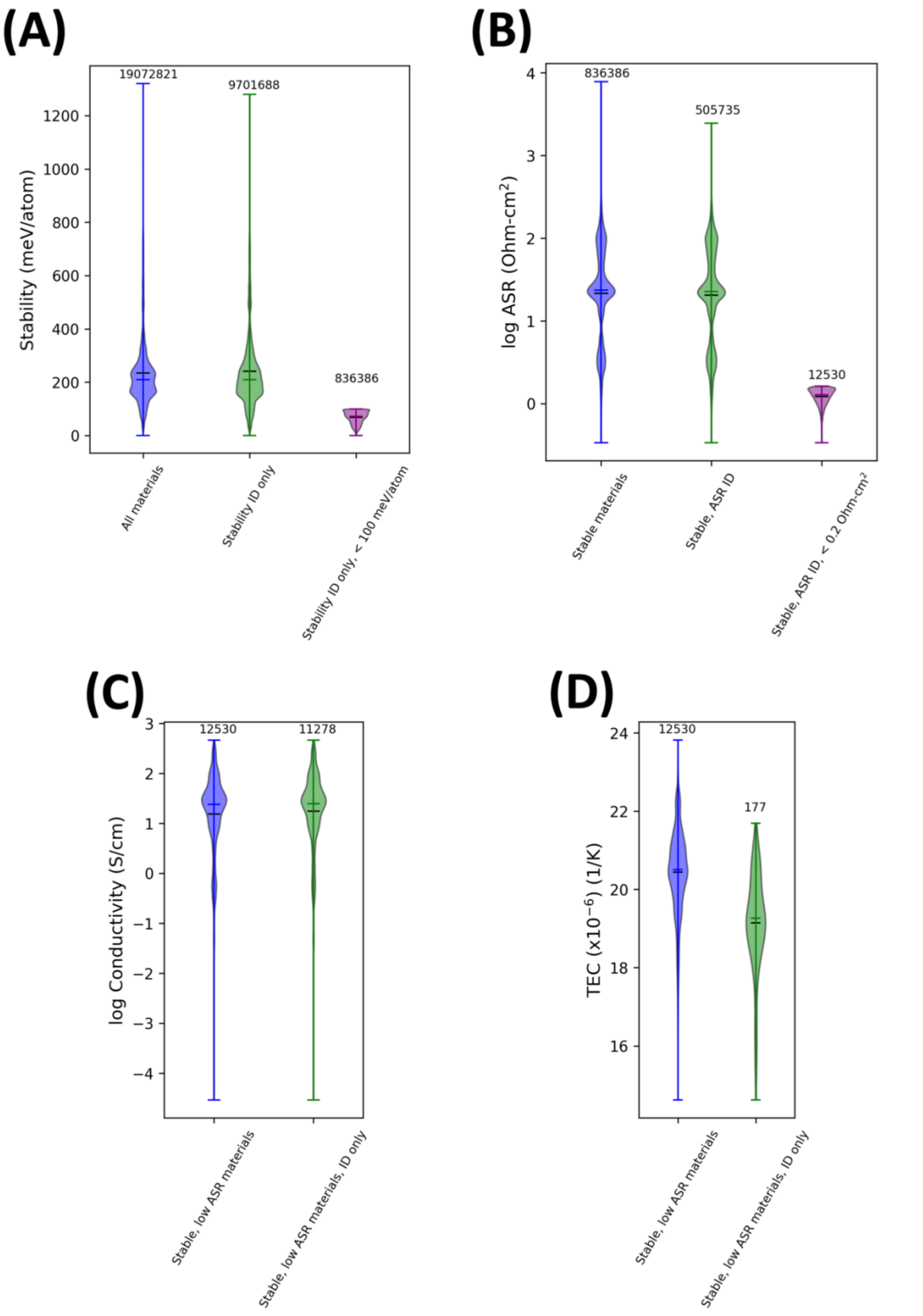

**Figure 8.** Violin screening plots showing distributions of materials property values at various stages of screening. (A) Stability distribution of all materials (blue), stability distribution of materials deemed in-domain by the reduced RMSE criterion (green), distribution of materials with in-domain stability predictions with stabilities less than 100 meV/atom (purple). (B) Distribution of ASR values for the subset of stable, in-domain materials from (A) (blue), distribution of stable materials with in-domain stability and ASR predictions (green), distribution of stable materials with low log ASR less than 0.2 Ohm-cm$^2$ (the value of BSCF, purple). (C) Distribution of conductivity values for the subset of stable, low ASR materials from (B) (blue), and the subset of those materials which have in-domain stability predictions (green). (D) Distribution of TEC values for the subset of stable, low ASR materials from (B) (blue), and the subset of those materials which have in-domain TEC predictions (green).

# 4. Summary

This study has addressed a critical need in the materials science community, namely the availability of persistent, easily accessible, and useable machine learning models of materials properties that can provide the user with predictions, uncertainties on those predictions (i.e., error bars), and guidance on domain of applicability. We have presented new approaches to error quantification and domain of applicability guidance, and shown that these approaches work well across a broad set of materials property datasets. Our unified approach to providing accurate property predictions, uncertainty estimates for those predictions, and guidance for model applicability domain is useful and important for informing improved multi-property materials discovery and design. The use of our approaches and the deployment of readily useable models has been made easier by our use of our MAST-ML code and the Garden-AI infrastructure. Given the ever-growing importance of machine learning approaches in materials science, and the increasing role of fast predictions of many materials properties in facilitating efficient and informed materials discovery and design choices, it is our hope that the construction of this set of 33 materials property models will inspire the community to develop new models of many more properties with informative error bars and domain guidance, host them using the Garden-AI infrastructure, and enable others to engage in state-of-the-art machine learning-informed materials research.

**Acknowledgements:**

The work of MAST-ML, error bars, domain and support for R. J., L. S., P.M.V. and D. M. were provided by the National Science Foundation under NSF Award Number 1931298 "Collaborative Research: Framework: Machine Learning Materials Innovation Infrastructure". The work on Foundry-ML and support for A. S., K. S. and B. B. were provided by the National Science Foundation under NSF Award Number 1931306 "Collaborative Research: Framework: Machine Learning Materials Innovation Infrastructure". The work on Garden-AI and support for O. P., W. E. and B. B. was provided by the National Science Foundation under NSF Award Number 2209892 "Garden: A FAIR Framework for Publishing and Applying AI Models for Translational Research in Science, Engineering, Education, and Industry".

**Data and Code Availability:**

All input and output data of the MAST-ML runs performed to evaluate the model fits, error bars, and domains of applicability are available online on Figshare (https://doi.org/10.6084/m9.figshare.26077015). All datasets used to fit the machine learning models are available on Figshare via the above link and are also hosted on Foundry-ML (https://www.foundry-ml.org).[84] All machine learning models are available online on the Garden-AI infrastructure (https://doi.org/10.26311/ep98-br79). Foundry-ML and Garden-AI leverage key software and service infrastructure developed by The Materials Data Facility[86,139] (https://www.materialsdatafacility.org) for dataset publication, and Globus (https://www.globus.org)[140] for data transfer, user authentication, search, curation flows, and more. An example Python notebook to invoke the Garden-hosted models is available on Figshare via the above link. The MAST-ML code is available on Github (https://github.com/uw-cmg/MAST-ML) and is installable via PyPi.